\documentclass{article} 
\usepackage[dvips]{graphicx}
\usepackage{amscd}
\usepackage{amsmath}
\usepackage{amsfonts}
\usepackage{amssymb}

\numberwithin{equation}{section}

\begin{document}
\title{\bf The Higgs Mass in the Substandard Theory \\
Talk at the 8th International Wigner Symposium in NYC, May 27, 2003 \\
{\small by} }
\author{E.L. Sch\"ucking \\
Department of Physics, New York University, NYC, NY 10003. \\
e-mail: elschucking@msn.com}
\date{}
\maketitle

\begin{abstract}
The Substandard theory deals with the standard model of
leptons, electro-weak gauge bosons and Higgs, excluding the chromodynamics of quarks.
The theory gives a geometric and algebraic interpretation of its $U(2)$
symmetry based on the Eguchi-Hanson metric
and predicts a Higgs mass $ m(H) = m(W) / \sqrt{sin \, \theta_{W}} = 115.3 GeV $.
Here $ m(W) $ is the mass of the charged gauge boson and $ \theta_{W} $ is the Weinberg angle. 
\end{abstract}

\eject

\section{Introduction}

The substandard theory goes back to a paper by an author whom Wigner
used to call \lq\lq my famous brother-in-law\rq\rq.
This paper by Jancsi\rq s second husband was published in
the {\it Proceedings of the Royal Irish Academy} in 1945 \cite{Dirac}.
I have never seen it quoted in the physics literature and
I discovered it through a reference in a book on the history of vector analysis
that made no use of it \cite{Crowe}.

Dirac took a pair of real quaternions $U$ and $V$ and formed their quotient $U/V$.
This got rid of a real number times a common phase factor
that for quaternions is a unit quaternion,
an element of the group $SU(2)$ for which he had no use in 1945.
At that time phases were thought to be an unphysical nuisance.
Instead of the homogeneous transformation 
\begin{equation}
\begin{pmatrix}
U' \\
V'
\end{pmatrix}
=
\begin{pmatrix}
a & b \\
c & d
\end{pmatrix}
\begin{pmatrix}
U \\
V
\end{pmatrix} \, ,
\qquad
{\mathbf a},{\mathbf b},{\mathbf c},{\mathbf d} \in {\mathbb H}
\end{equation}
he used the broken linear transformation which represented Lorentz transformations and more.
 
The mathematician Lambek \cite{Lambeck} remembered:
\lq\lq I recall telling Dirac in 1949 that
I could derive his equation with the help of quaternions.
After thinking quietly for several minutes, as was his habit before speaking,
he said \lq Unless you can do it with real quaternions, I am not interested\rq.\rq\rq
 
For unit determinant Dirac's homogeneous group is isomorphic to
$\text{Spin}^{\uparrow}(1,5)$,
the cover of the orthochronic subgroup of $SO(1,5)$ that contains the Lorentz group and a $U(1)$.
The pair of quaternions $U$ and $V$ can be interpreted
as containing left-handed electrons and neutrinos.
By multiplying this pinor with a unit quaternion from the right (the enhanced group)
we have the fundamental representation of left-handed leptons
under the combination of the Lorentz group times $U(2)$.
A real quaternion can be described here as a pair of complex numbers by writing 
\begin{equation}
U = x + y \, {\mathbf j}
\end{equation}
with the complex numbers x and y and \lq\lq ${\mathbf j}$\rq\rq\
as one of the imaginary quaternionic units.
I might also mention here that Dirac's group gets us two additional spatial dimensions
that can be used as an internal symmetry space.
A reflection in this 2-space (charge conjugation)
combined with a reflection in the 3-space (parity)
is continuously connected with the identity and might thus, as $CP$,
survive the split of the five-dimensional space into its external and internal parts. 

\section{Quaternions}

The great importance of quaternions lies in their uniqueness
first proven by Frobenius in 1877 through the theorem \cite{Palais} 
\lq\lq Any real finite associative division algebra is either
${\mathbb R}$, ${\mathbb C}$, or ${\mathbb H}$.\rq\rq\
Since ${\mathbb R}$ is of \lq unreasonable effectiveness\rq\ (Wigner \cite{Wigner})
in physics one might hope that ${\mathbb C}$ and ${\mathbb H}$ are too.

From Kelvin \cite{Ebbinghaus} to Gell-Mann \cite{Gell-Mann}
quaternions have not always found friends in physics.
Lord Kelvin stated:
\lq\lq Quaternions came from Hamilton after his really good work had been done;
and though beautifully ingenious,
have been an unmixed evil to those who have touched them in any way.\rq\rq
 
But he also stated: \lq\lq vector is a useless survival, or off­shoot from quaternions,
and has never been the slightest use to any creature.\rq\rq
 
With the advent of special relativity and later,
with Dirac's equation, quaternions became fashionable again,
but not the real kind.
People used bi-quaternions, {\it id est}, pairs of quaternions $ a + b {\mathbf e} $
where $ {\mathbf e} $ commutes with the quaternions $a$ and $b$ and
gives ${\mathbf e} \ {\mathbf e} = -1$.
These bi-quaternions are equivalent to $ 2 \times 2 $ complex matrices,
have divisors of zero like
\begin{equation}
(i + {\mathbf e})(i - {\mathbf e}) = 0,
\end{equation}
and they are just another mathematically uninteresting algebra.
 
Real quaternions do play a basic and crucial role in physics
because the Clifford algebra $Cl(1,3)$ of the real Lorentz metric
\begin{equation}
ds^{2} = \, \bigr(dx^{0}\bigr)^{2}
- \, \bigr(dx^{1}\bigr)^{2}
- \, \bigr(dx^{2}\bigr)^{2}
- \, \bigr(dx^{3}\bigr)^{2} 
\end{equation}
is quaternionic \cite{Budinich}.
To describe spinors four complex matrices $ \gamma^{\mu} $ are introduced with
\begin{equation}
\gamma^{\mu} \gamma^{\nu} \, + \, \gamma^{\nu} \gamma^{\mu} \, = \, 2 \, \eta^{\mu\nu} \, 1.
\end{equation}
The Clifford algebra is then generated by 1 and the products of the $ \gamma^{\mu} $.
But that is the wrong one.
It belongs to a complex metric with complex $ dx^{µ} $
that do not know the difference between space and time.
Metrics have to be real or Hermitean.
Complex metrics make no sense physically.
The complex Clifford algebra contains a lot of physical junk.
What we have to consider is the
\underbar{real} Clifford algebra that belongs to the \underbar{real} Minkowski metric.

The Clifford algebra $Cl(1,3)$ consists of $ 2 \times 2 $ quaternionic matrices
and can be generated by the matrices
\begin{equation}
\gamma^{0} \, = \,
\begin{pmatrix}
0 & j \\
-j & 0
\end{pmatrix} , \
\gamma^{1} \, = \, 
\begin{pmatrix}
-j & 0 \\
0 & j
\end{pmatrix}, \
\gamma^{2} \, = \,
\begin{pmatrix}
k & 0 \\
0 & k
\end{pmatrix}, \
\gamma^{3} \, = \,
\begin{pmatrix}
0 & j \\
j & 0
\end{pmatrix}
\end{equation}
leading us back to Dirac's group $\text{Spin}^{\uparrow}(1,5)$.

If we were to take the opposite signature for the Minkowski metric
we get the Clifford algebra $Cl(3,1)$
that is the algebra $R(4)$ of real $ 4 \times 4 $ matrices.
This algebra can be generated by a Majorana representation with real
\begin{equation}
\gamma^{0} \, = \,
\begin{pmatrix}
0 & i\sigma_{2} \\
i\sigma_{2} & 0
\end{pmatrix} , \
\gamma^{1} \, = \, 
\begin{pmatrix}
0 & \sigma_{1} \\
\sigma_{1} & 0
\end{pmatrix}, \
\gamma^{2} \, = \,
\begin{pmatrix}
1 & 0 \\
0 & -1
\end{pmatrix}, \
\gamma^{3} \, = \,
\begin{pmatrix}
0 & \sigma_{3} \\
\sigma_{3} & 0
\end{pmatrix}
\end{equation}
and Pauli matrices ($\sigma_{r} = 1, 2, 3$).
 
If one represents quaternions by complex $ 2 \times 2 $ matrices
the transition from one signature to the other is easily achieved through the complex.
Thus, the difference between the two signatures has, historically, been deemed irrelevant.
 
However, the difference is profound and fundamental
if one is interested in the question of the origin of the imaginary in physics
and tries to enlarge space to accommodate internal symmetries.
The quaternions can provide us with an imaginary unit that is indispensable for quantum mechanics.
Moreover, the fundamental representation of the Clfford algebra $Cl(3,1)$
leads to a real 4-component Majorana spinor describing chargeless particles
that have yet to find a home in the particle data tables.
Further, if we add two internal symmetry dimensions to spacetime
the Clifford algebras $Cl(1,5)$ and $Cl(5,1)$ are both quaternionic.

For both signatures we get a chiral imaginary unit by the pseudoscalar
\begin{equation}
\gamma^{5} \, = \, \gamma^{0} \, \gamma^{1} \, \gamma^{2} \, \gamma^{3} \, , \qquad
\bigr(\gamma^{5}\bigr)^{2} = -1 \, .
\end{equation} 
Only for $Cl(1,3)$ can we choose this $\gamma^{5}$ as $i$ times the unit matrix
without enlarging the ground field where $\mathbf i$ is now a vectorial unit quaternion
which becomes the generator of the weak hypercharge transformations. 

\section{The Algebraic Conjecture}

The ideas of Gauss, Riemann, Levi-Civita and Hessenberg
suggested making the local vector structure in the tangent space of a manifold path-dependent.
Einstein interpreted then this connection as a gravitational field.
However, length was kept path-independent despite Weyl's wishes,
and, obvious and self-evident, the imaginary \lq\lq $i$\rq\rq\ in a wave function
was considered a universal fixture and the same everywhere and certainly not path-dependent.
The same would be assumed for a quaternionic structure.

I find it tempting to consider the complex structure
and the quaternionic structure of algebra as space-time and path-dependent.
The reason for this is as follows:
the minimal Higgs field whose existence is widely assumed is a scalar quaternionic field.
The automorphism group of the quaternions is $SO(3)$.
To bring in a compatible complex structure we have to extend it
into the group $Spin^{c}(3)$ that is isomorphic to $U(2)$ containing the additional $U(1)$.
This provides us with an algebraic rationale for the existence of the electro-weak group $U(2)$.
Gauging then accounts for the four electro-weak gauge bosons. 
The usual representation of quaternions in terms of complex numbers
is given as real multiples of $SU(2)$ matrices.
The three imaginary quaternionic units are taken as minus $i$ times the Pauli matrices
\begin{equation}
{\mathbf i} \longrightarrow \frac{1}{i} \sigma_{1} \, , \qquad
{\mathbf j} \longrightarrow \frac{1}{i} \sigma_{2} \, , \qquad
{\mathbf k} \longrightarrow \frac{1}{i} \sigma_{3} \, .
\end{equation} 
The Higgs field transforms then as
\begin{equation}
\begin{pmatrix}
{\phi}^{\prime}_{1} & - {\phi}^{\prime *}_{0} \\
{\phi}^{\prime}_{0} & {\phi}^{\prime *}_{1}
\end{pmatrix}
=
e^{\frac{i}{2} {\tilde \lambda} \cdot {\tilde \sigma} }
\begin{pmatrix}
\phi_{1} & - {\phi}^{*}_{0} \\
\phi_{0} & {\phi}^{*}_{1}
\end{pmatrix}
\begin{pmatrix}
e^{\frac{i}{2}f} & 0 \\
0               & e^{-\frac{i}{2}f}
\end{pmatrix}
\end{equation}
and is usually assumed to be a multiple of the quaternion $\mathbf j$.
We turn now to the geometric picture. 

\section{The Geometric Conjecture}

If one accepts the presence of a $U(2)$ internal symmetry group
one may raise the question of its action in an internal symmetry space
and thus about the nature of the internal symmetry space.
Assuming with Theodor Kaluza a Riemannian metric for the internal symmetry space
one might think that at least three dimensions are needed to accommodate $U(2)$
as an isometry group in the internal space.
However one can save dimensions by realizing the group as
a dynamical group in the cotangent bundle of a manifold,
as known from the symmetry of the Kepler problem.
In this case one can get away with two dimensions
if one realizes $U(2)$ on the cotangent bundle of a two-dimensional sphere.
While the rotations of a sphere use 3 parameters
we get a fourth one now by rotating the tangent plane
about its osculating point while keeping the sphere fixed.
In fact, this appears to be the simplest choice.
The metric of this four-dimensional manifold becomes fixed
if we make the far-reaching assumption that
we want the manifold to be Ricci-flat and K\"ahler.
The Ricci-flatness is suggested as analogy to Einstein's vacuum field equations
while the K\"ahler property brings in complex and symplectic structures.
While the complex structure is desirable,
since the Higgs is a complex representation of a quaternionic structure,
the best excuse for the introduction of a symplectic structure
is the hope that it might be of help for the quantization.
It turns out that the only Ricci-flat $U(2)$-invariant K\"ahler manifold 
is the cotangent bundle of an $ S^{2} $ and depends only on the radius of the sphere.
Tohru Eguchi and Andrew Hanson discovered this space in 1978 \cite{Eguchi}. 

\section{The Eguchi-Hanson Space}

The metric of the Eguchi-Hanson space is given by
\begin{subequations}
\begin{gather}
ds^{2} = \, 2 \, \frac{\sqrt{a^{2}+R^{2}}}{R}
\left[ \,
d\phi_{\alpha} \, d\phi^{*}_{\alpha}
\, - \,
a^{2} \frac{\vert \phi_{\alpha} \, d\phi^{*}_{\alpha} \vert^{2}}{R(a^{2}+R^{2})}
\right] \, , \\
R \, = \, \vert \phi_{1} \vert^{2} \, +  \, \vert \phi_{0} \vert^{2} \, , \quad
a \, = \, const. \, , \quad \phi_{\alpha} \in {\mathbb C}^{2} .
\end{gather}
\end{subequations}
Here $R$ is quadratic in the complex coordinates.
The $U(2)$ group acts on the space as follows:
\begin{equation}
\begin{pmatrix}
{\phi}^{\prime}_{1} & - {\phi}^{\prime *}_{0} \\
{\phi}^{\prime}_{0} & {\phi}^{\prime *}_{1}
\end{pmatrix}
=
e^{\frac{i}{2} {\tilde \lambda} \cdot {\tilde \sigma} }
\begin{pmatrix}
\phi_{1} & - {\phi}^{*}_{0} \\
\phi_{0} & {\phi}^{*}_{1}
\end{pmatrix}
\begin{pmatrix}
e^{\frac{i}{2}f} & 0 \\
0               & e^{-\frac{i}{2}f}
\end{pmatrix}
\end{equation}
It is clear from the metric that the space becomes flat for infinite $R$
and that the hypersurfaces $ R = const $ are the orbits of the group $U(2)$.
They are homogeneous spaces which are deformed three-dimensional spheres
in which opposite points have been identified.
To get a better idea of these spaces imagine an $ S^{3} $
with three orthogonal geodesics through a point.
Applying now left-translations to these geodesics
we get a threefold set of fibers through every point
converting this space into a continuous jungle gym.
While the original right- and left-invariant metric of the $ S^{3} $
was given by
$$
- \, ds^{2} = \, (\omega_{1})^{2} \, + \, (\omega_{2})^{2} \, + \, (\omega_{3})^{2}
$$
in terms of the three differential forms, the deformed space has the metric
$$
- \, ds^{2} = \, (\omega_{1})^{2} \, + \, (\omega_{2})^{2}
\, + \, \lambda^{2}(\omega_{3})^{2}
$$
where a positive factor $\lambda$ describes the stretching or compression along the 3-axis.
The metric still remains invariant under all right translations
and under left translations in direction of the 3-axis.
The space thus remains homogeneous and in each point axially symmetric.
Some relativists know these spaces under the name Bianchi type IX \cite{Bianchi}.
Istv\'an Ozsv\'ath, the father of one of the previous speakers,
devoted a good deal of his time to embed (not like journalists)
these spaces isometrically in higher-dimensional Euclidean spaces
to look at them from outside.
He called them Dantes \cite{Ozsvath}
after the poet who conceived Earth and Heaven as the two parts of an $ S^{3} $
with the devil at the center of the Earth as the antipode of you know who. 
While I tried to entertain you with my chatter
it must have dawned on you that there is something fundamentally wrong
with the Eguchi-Hanson metric.
It cannot be denied:
this metric has a singularity for $ R = 0 $
unless the constant $a$ is equal to zero turning Eguchi-Hanson into a flat space.
However, not all is lost.
By blowing up, a favorite activity of algebraic geometers,
one finds that one can save the situation by identifying opposite points
on the hypersurfaces $ R = const $ (identifying $ \phi_{\alpha} $ by $ -\phi_{\alpha} $)
and inserting at $ R = 0 $ a two-sphere of radius $ \sqrt{a/2} $. 
Here we do finally meet the sphere whose cotangent bundle is the Eguchi-Hanson space. 

\section{The Gauge Bosons}

Since the group $U(2)$ is four-dimensional but 
has three-dimensional orbits in the Eguchi-Hanson space
its four infinitesimal generators known as Killing vectors
are tangent to the hypersurfaces $R = const$.
On these hypersurfaces the group acts multiply transitive.
In each given point one of the three Killing vectors will vanish
and thus correspond to an infinitesimal rotation
while the other three are the generators of translations.
The four Killing vectors are independent with respect to constant coefficients but not in a point.

The gauge bosons are understood as the space-time extensions
of the differential forms in the Eguchi-Hanson space.
The components there correspond to the so-called ghost fields.
We have then that translational Killing vectors measure masses while rotational ones measure charges.
In the standard normalization where the Higgs has only one real component
(in quaternionic language it is proportional to $\mathbf j$),
we are dealing with the tangent space at the North pole of the $S^{2}$,
the Killing vector that vanishes there is the generator of the electric charge.
In this case the corresponding gauge boson will be massless.
From the length of the other Killing vectors we obtain the ratio of the masses of the gauge bosons. Clearly, the ratio of the Killing vectors are given by the stretching factor
and this gives us now a geometrical interpretation of the Weinberg angle.
We have 
\begin{equation}
\text{cos} \, \theta_{W} \, = \, \frac{R}{\sqrt{a^{2}+R^{2}}} \, = \, \frac{m(W)}{m(Z)} \, .
\end{equation}
Because of homogeneity the Weinberg angle depends only on $R$. 

\section{The Higgs}

The Higgs as seen here is the phase space of a two-sphere.
Two-spheres are conveniently thought embedded into a Euclidean $R^{3}$
with coordinates $x$, $y$, and $z$.
The Higgs defines a two-sphere through a Hopf map 
\begin{equation}
x \, + \, i \, y \, = \, 2 \, \phi_{0} \, \phi^{*}_{1} \, , \qquad
z \, = \, \vert \phi_{0} \vert^{2} \, - \, \vert \phi_{1} \vert^{2} \, .
\end{equation}
This has the result that for constant $R$ the radius of this two-sphere is also constant 
\begin{equation}
x^{2} \, + \, y^{2} \, + \, z^{2} 
\, = \,
\bigr( \vert \phi_{0} \vert^{2} \, + \, \vert \phi_{1} \vert^{2} \bigr)^{2} \, .
\end{equation}
The only mass-scale available in the Eguchi-Hanson space
is the inverse radius of the two-sphere for $R = 0$.
It is suggestive to identify this mass with the Higgs mass 
\begin{equation}
m(H) \, = \, \sqrt{\frac{2}{a}} \, .
\end{equation}
However, since we are looking for a characteristic length of a sphere
a factor like $2$ or $\pi$ could not be excluded from guessing.
We have then from the Killing vectors 
\begin{equation}
m(H) \, = \, m(Z) \, \frac{\text{cos}\, \theta_{W}}{\sqrt{\text{sin}\, \theta_{W}}}
\end{equation}
giving \cite{Hagiwara}
\begin{equation}
m(H) = 115.3 \, \text{GeV}.
\end{equation}
We can introduce now a potential energy for the Higgs field
into the 4-dimensional Lagrangean with a Lagrangean multiplier $\lambda$ by writing 
\begin{equation}
V(\vert\Phi\vert) \, = \, \lambda
\left( \, \vert\Phi\vert^{2} \, - \, \frac{1}{\lambda a} \, \right)^{2} 
\, = \,
\lambda \, \vert\Phi\vert^{4} \, - \, \Phi^{2} \, \mu^{2} \, + \, \frac{1}{\lambda a^{2}}
\end{equation}
which gives 
\begin{equation}
m(H) \, = \, \sqrt{\frac{2}{a}} \, = \, \sqrt{2\mu^{2}} \, .
\end{equation}
This expresses the condition that the internal symmetry space remains a two-sphere. 

\section{Conclusion}

The shortcomings of the substandard theory are already addressed by its name
but a number of features of the standard model appear to fall into place when viewed from this angle. 
I am grateful to Jerome Epstein for discussions and to Jie Zhao for work on the manuscript.

\end{document}